\documentclass[twoside]{article}
\usepackage{fleqn,espcrc2}

\usepackage{graphicx}
\usepackage[figuresright]{rotating}

\newcommand{\AmS}{{\protect\the\textfont2
  A\kern-.1667em\lower.5ex\hbox{M}\kern-.125emS}}

\title{Topologically non-trivial configurations in 3-dimensional Yang-Mills theory}

\author{Pushan Majumdar\address[IMSC]{Institute of Mathematical Sciences,
	C.I.T. Campus, Taramani, Chennai - 600113, India.}%
        \thanks{Present address: Dept. of Theoretical physics, Tata 
	Institute of Fundamental Research, Homi Bhabha Road, Mumbai 400005. India.}
        and
        Dong-Shin Shin\addressmark[IMSC]}
       
\begin{document}

\begin{abstract}
Recently Anishetty, Majumdar and Sharatchandra have proposed a way of characterizing
topologically non-trivial configurations for 2+1-dimensional Yang-Mills theory in a
local and manifestly gauge invariant manner. In this paper we develop criteria to locate
such objects in lattice gauge theory and find them in numerical simulations.
\vspace{1pc}
\end{abstract}

\maketitle

\section{Introduction}
Monopoles are expected to play an important role in confining quarks in QCD. In lattice
simulations, one usually looks for U(1) monopoles by fixing an Abelian gauge \cite{2}.
However it is important to have a gauge invariant way of detecting monopoles in 
non-Abelian gauge theories. 
In a recent paper \cite{1} Anishetty, Majumdar and Sharatchandra have
given a criterion for
characterizing topologically non-trivial configurations in 3-dimensional
SU(2) Yang-Mills theory. This criterion is local and manifestly gauge invariant.
This was achieved by formulating the theory in terms of 
gauge invariant variables closely related to gravity. 
This rewriting also turns out to be a duality transformation as it neatly separates the
``spin waves" from the ``topological degrees of freedom". We see this explicitly from
simulation data (see fig.1). 
Reformulation of the theory, even if it gives us the criterion, does not tell 
us whether such configurations actually occur. That is an important dynamical question
and we attempt to answer it by simulating 2+1-dimensional SU(2) lattice gauge theory and 
checking for the topologically non-trivial configurations. From now on, motivated by
the fact that finally we want to look at the 3+1-dimensional theory, we will call these 
topologically non-trivial objects in three dimensions as monopoles, even though they
are actually instantons of the 3-dimensional theory.

\section{Method}
The partition function of 3-dimensional Yang-Mills theory is
\begin{equation}
Z\!=\!\int\;{\cal D}A_i^a(x)\;exp\left ( -\frac{1}{2\kappa^2}\int\;d^3 x
F_{ij}^a(x)F_{ij}^a(x)\right ) 
\end{equation}
where $i,j$ run over $1,2,3$ and 
\begin{equation}
F_{ij}^a(x)=\partial_i A_j^a-\partial_j A_i^a+\epsilon^{abc}A_i^b A_j^c
\end{equation}
is the usual field strength.

To identify the monopoles of the theory one can use the 
orthogonal set of eigenfunctions of a positive symmetric matrix.
For that purpose we consider the eigenvalue equation of the matrix
$F_{ij}^a(x)F_{kj}^a(x)
=I_{ik}(x)$ for each space-time point $x$.
\begin{equation}
I_{ik}(x)\chi_k^A(x)=\lambda^A(x)\chi_i^A(x)
\end{equation}
Here $A$ is not summed over but labels the eigenvalues.

Isolated points where $I_{ik}$ have triply degenerate eigenvalues are special, and 
have topological significance. At such points, the vector fields formed by the
eigenvectors of $I_{ij}$ are singular. The index of the vector field at the singular 
point is the monopole number.
Thus the monopoles in any Yang-Mills configuration $A_i^a(x)$ can be located in
terms of the eigenvectors $\chi_i^A(x)$.
One can also construct coordinate system using $\chi^A(x)$. Integral curves of this
vector field are equivalent to the $r$-coordinate. In that case, the coordinate
singularities of this coordinate system correspond to the monopoles.

For our simulations, we choose the usual Wilson action for SU(2) gauge theory
\begin{equation}
 S_W=\frac{\beta}{2} \sum_{plaquettes} tr (U_{ij}U_{jk}U_{kl}U_{li})
\end{equation}
and measure the basic plaquette at every site. The plaquette variable 
can be written as 
\begin{equation}
exp\,i\,F_{ij}^a\sigma^a = cos (|F_{ij}|) + i{\hat F}_{ij}^a\sigma^a\, sin(|F_{ij}|)
\end{equation}
where ${\hat F}_{ij}^a$ is the unit vector corresponding to $F_{ij}^a$.
Therefore $F_{ij}^a={\hat F}_{ij}^a\, cos^{-1} (cos(|F_{ij}|))$. Once we get $F_{ij}^a$,
we construct $I_{ij}$ as $I_{ij}=F_{ik}^aF_{jk}^a$.

At the location of the monopole, all three eigenvalues of $I_{ij}$ should be 
degenerate. However
on the lattice we do not expect the eigenvalues to become exactly degenerate, but we 
look for sites where the difference between the eigenvalues are less than some
small but non-zero number. Henceforth we shall refer to this number as cut-off. For
spherically symmetric monopoles
in continuum, around the location of the monopole, one of the eigenvalues 
will become non-degenerate with the other two which would still be 
degenerate. The eigenvector corresponding to this eigenvalue will show a radial 
behaviour. On the lattice we choose the eigenvector corresponding to 
the largest eigenvalue and plot it at the site of the monopole and its nearest 
neighbors to check for this radial behavior.

\section{Results}
In our simulation, we look at various lattice sizes and couplings $\beta$. In three 
dimensions since $\beta$ has dimension of length (to leading order), we keep the ratio
between $\beta $ and the lattice size fixed. This increases the equilibriation time for
the 
larger lattices. For lattice size $64^3$ (the largest lattice we consider) the
equilibriation time is of the order of $240$
updates. We ignore the first $300$ updates and after that take measurements in every 
successive update for $300$ updates.

The number of monopoles is very sensitive to the choice of the cut-off. A small variation
in the cut-off can change the number of monopoles detected by an order of magnitude.
To choose the cutoff we look at the distribution 
of the smallest eigenvalues for the various lattice sizes. Then we choose the cut-off
to be half the mean minimum eigenvalue for each lattice size. This minimizes the chance
that the eigenvalues become degenerate purely due to statistical effects.
The various values for which we take data are shown in Table \ref{table:I}.
\begin{table}[htb]
\caption{ The choice of cut-off for various $\beta$}
\label{table:I}
\begin{tabular}{@{}cccc}
\hline \\
$\beta$ & lattice size & mean lowest & cut-off \\
&&eigenvalue& \\
\hline \\
1.5 & 16 & 0.253 & 0.1265 \\
2.25 & 24 & 0.129 & 0.0646 \\
3 & 32 & 0.079 & 0.0397 \\
3.75 & 40 & 0.0521 & 0.026 \\
4.5 & 48 & 0.0368 & 0.0184 \\
5.25 & 56 & 0.0264 & 0.0132 \\
6 & 64 & 0.0217 & 0.0109 \\
\hline
\end{tabular}
\end{table}
With these parameters, typically we find one lattice site in every two or three 
measurements which has the difference of eigenvalues less than the cut-off. 
For a few configurations (roughly 3 or 4 out of the 300 probed), in every lattice size,
we find more than one site in a single configuration which meets the eigenvalue
criterion. 
After this we look at the eigenvectors corresponding to the largest eigenvalue around the 
lattice site which satisfies the criterion for the degenerate eigenvalues. Among 
them we look for sites that have at least three non-coplanar eigenvectors which converge 
to or diverge from a point. Figure 2 shows a typical configuration we are looking for.
In order to make sure that the eigenvectors really converge, we rotate the
eigenvectors and check that they remain
convergent from all angles. In our formulation, the eigenvectors only specify the rays 
and not the direction of the vector. So we do not distinguish between monopoles and 
antimonopoles. Our results are shown in Table \ref{table:II}.

\begin{table}[htb]
\caption{Total number of sites where the eigenvalues are degenerate and around which the 
eigenvectors show radial behavior.}
\label{table:II}
\begin{tabular}{cccc}
\hline \\
$\beta$ & lattice size & degenerate & radial behavior \\
&& eigenvalues & of eigenvector \\
\hline \\
1.5 & 16 & 59 & 22  \\
2.25 & 24 & 58 & 12  \\
3& 32 & 67 & 6  \\
3.75 & 40 & 63 & 14 \\
4.5 & 48 & 78 & 16 \\
5.25 & 56 & 52 & 7 \\
6 & 64 & 69 & 18 \\
\hline
\end{tabular}\\[8pt]
Note that the sites which show radial behavior also have degenerate eigenvalues. 
\end{table}

\section{Conclusions}
We have seen that it is indeed possible to detect topologically non-trivial
configurations using the criterion presented in \cite{1}. Our data also indicates 
that configurations extended over many cells or plaquettes are favored compared to the
ones over a single cells or plaquettes. Moreover only a fraction of the monopoles
detected have a spherical symmetry.

\begin{figure}[htb]
\includegraphics[width=18pc]{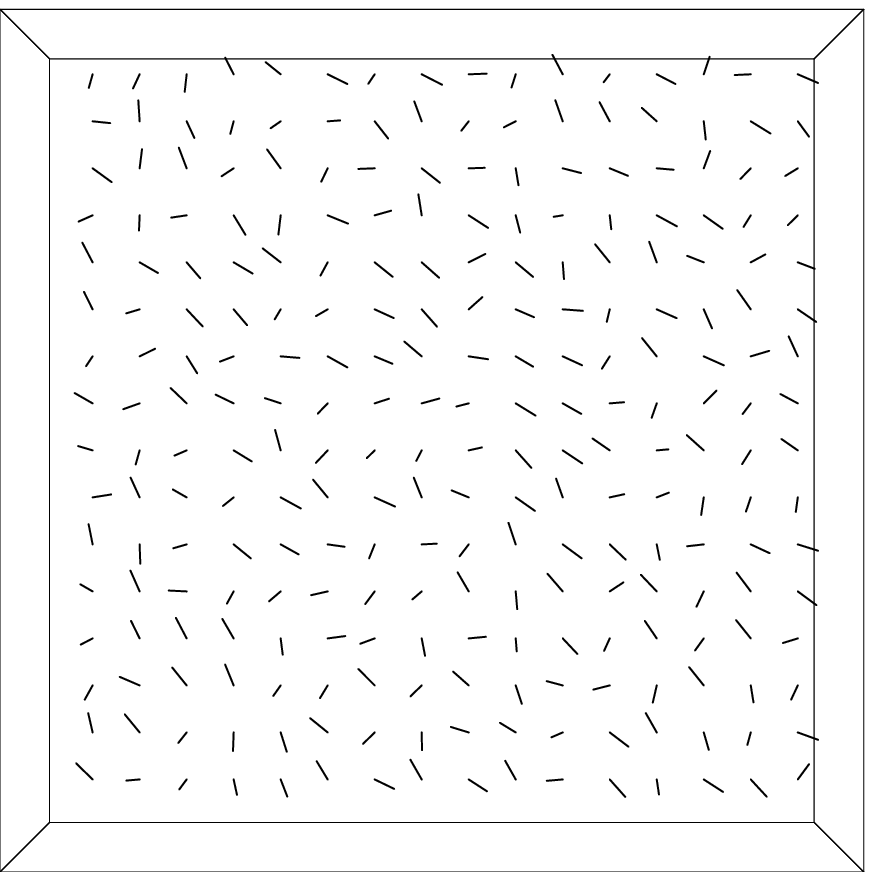}
\caption{ Snapshot of a x-y plane with eigenvectors projected to the plane.
Lattice size 16.}
\end{figure}

\begin{figure}[htb] 
\includegraphics[width=15pc]{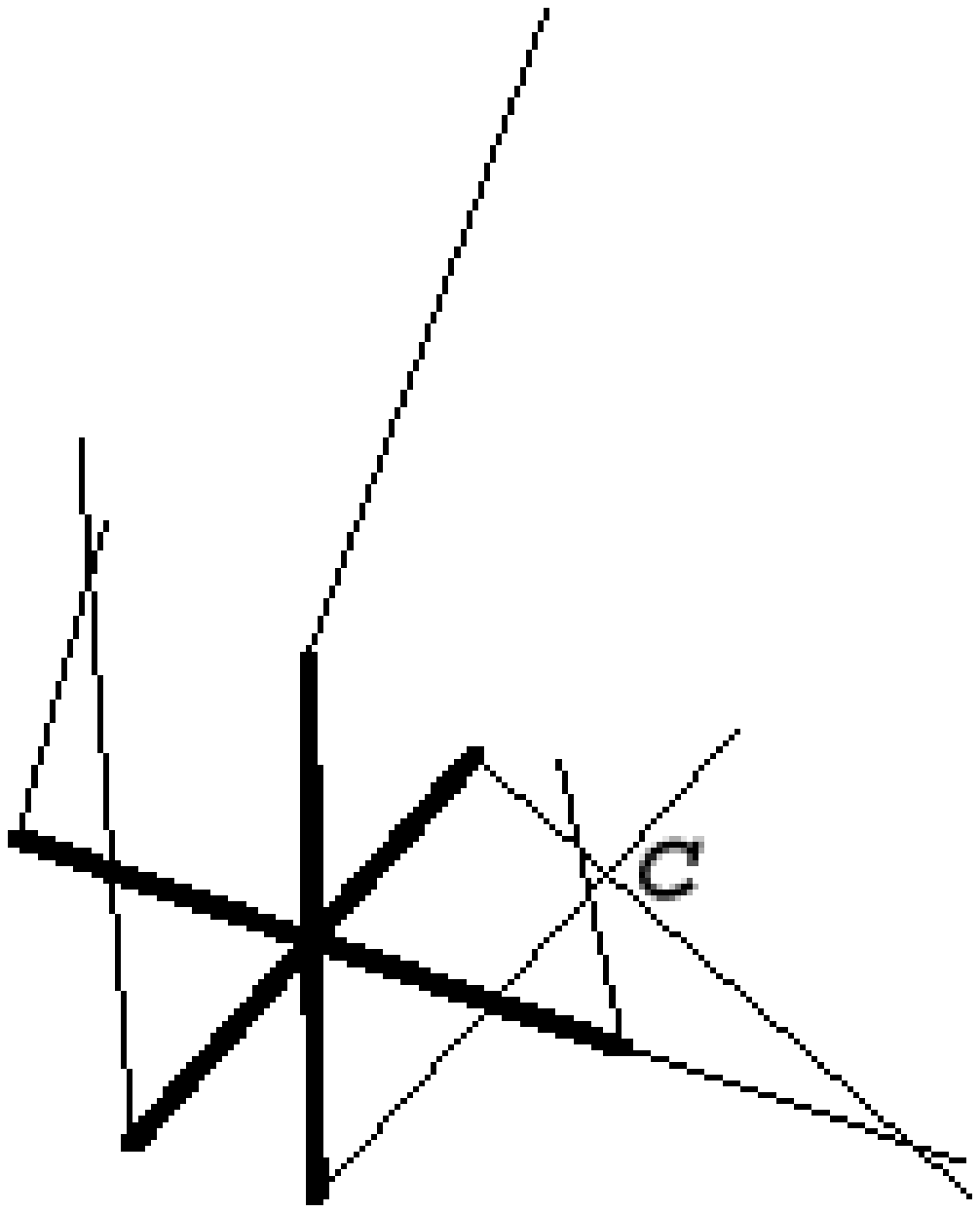}
\caption{ Eigenvector configuration
we are looking for. The eigenvectors shown here are scaled to twice their size to
show their crossing explicitly. Lattice size is 56.}   
\end{figure}
 
\end{document}